# Faculty Experience with Generative AI Integration into the Teaching and Learning Process

Brent A. Anders, PhD., American University of Armenia, ORCiD: 0000-0002-0822-8224

Anait Akopyan, M.A., PhD(c), American University of Armenia, ORCiD: 0000-0003-2793-4463

## Abstract

This study was conducted to enhance our knowledge and understanding of the lived experiences of faculty regarding their experiences with generative artificial intelligence (AI) in teaching and learning within the English and Communication Program at the American University of Armenia. This phenomenological qualitative research study used face-to-face, in-depth, open-ended interviews with five faculty members from the program. Faculty expressed different levels of confliction, voicing both positive and negative issues/concerns regarding their experiences with generative AI in the teaching and learning process. Five categories and 11 themes emerged from the analysis of these interviews, such as concerns about academic integrity and student learning, and a lack of faculty AI training, understanding, and/or experience. Emerged themes indicated that faculty need additional support, to include training and leadership guidance, in order to better perceive the importance and benefits of effective AI utilization and AI literacy development. Faculty lacked full experience with and understanding of generative AI capabilities or ongoing developments. Overall faculties' perceptions and attitudes of generative AI within teaching and learning can be characterized as a mix of reservation and optimism.

## Plain Language Summary

Study exploring how university faculty experienced generative AI in the teaching and learning process. In-depth interviews revealed mixed feelings, with concerns about academic integrity and the need for more AI literacy training. Faculty generally feel both cautious and hopeful about using AI in education.

## Introduction

Although signs and articles were present before the release of the popular generative AI: ChatGPT, by OpenAI, most faculty were not prepared for a world of free access to such a powerful technological tool (Anders, 2022; Meyer et al., 2023; Xie et al., 2021; Zhang & Aslan, 2021). As AI technologies are being integrated in different ways and within different parts of academia, there is an important need to fully understand the ways that educators are using and reacting to this new technology. The revolutionary nature of generative AI holds the potential of transforming higher education for the better. Yet this can only be successfully achieved by understanding the lived experiences of what faculty go through with AI in the teaching and learning process. Gaining these greater inputs will allow everyone in academia to make better informed decisions regarding generative AI utilization and implementation. By thoroughly understanding faculty's phenomenological experiences, challenges, and successes, we will contribute important and useable insights that can guide faculty as well as educational institutions in optimizing the integration of AI technologies into teaching and learning.

This research seeks to assist in better bridging the gap between the rapid advancements in technology and current educational practices. The real contextual factors that shape how AI technologies can best be used in education must be fully understood; phenomenology, as a research framework, provides an in-depth and holistic approach to obtain real subjective faculty experiences. Through this methodology we obtained a greater understanding of how generative AI is used and experienced in instructional design, pedagogical approaches, help seeking behaviors, and the overall teaching and learning experience. The following research questions were used to help achieve study findings:

> **RQ1** What are the perceived experiences of faculty with their use of generative AI in the teaching and learning process?
>
> **RQ2** How do faculty describe their motivations and/or hesitations (worries/concerns) about using generative AI in teaching and learning?
>
> **RQ3** How do faculty describe any impediments in the use of generative AI in the teaching and learning process?

### ***Background***

Due to the relatively new nature and availability of AI to the general public and academia, there has not been a lot of phenomenological research in this field. Although many use-of-AI-in-education surveys have been conducted (Barnes & Tour, 2026; Kaplan-Rakowski et al., 2023;), these have offered general quantitative informational results without the deeper meaning as to the "why" that comes from qualitative research in general and phenomenological research specifically. However, the research by Barnes, Tour, Kaplan-Rakowski, and his fellow researchers from Texas, USA and Canada, have provided useful information for other research, like ours, to build upon. Their study suggests that "regardless of their exact perceptions, teachers identified both advantages and challenges regarding GAI [generative artificial intelligence] integration in education" (Kaplan-Rakowski et al., 2023, p. 329), with the overall recommendation that specific training and professional development is needed to help teachers build trust in the new technology.

Other researchers, such as Giovanni Zimotti, Claire Frances, and Luke Whitaker, from the University of Iowa, United States (2024), conducted a survey on second language (L2) instructors and obtained the following:

> …experience with ChatGPT was found to significantly influence instructors' levels of excitement. Those who had previously engaged with the tool expressed greater enthusiasm for its wider application compared to their counterparts with no such experience. This suggests that familiarity may mitigate apprehension and foster a more optimistic outlook on the potential of ChatGPT in educational settings… (2024, p. 18)

Their recommendation was that "we must quickly learn to use them [AI tools], understand the challenges and benefits, and develop both instructional approaches and policies…. we need to be agile and adaptable when faced with rapidly changing instructional conditions" (p. 16). These finding correlate with survey results on Middle Eastern/North African faculty, which identified "a positive correlation between knowledge about AI and attitudes towards AI as well as a significant positive correlation between AI knowledge and overall KAP [Knowledge, Attitudes, and Practices] scores" (Hasan et al., 2024, p. 2). They recommend that faculty engage with AI and attend training through workshops, conferences, as well as other online resources to be informed and "to capitalize on AI's benefits while navigating its challenges" (p. 21).

There have been fewer phenomenological studies conducted dealing with generative AI. Dubois du Bellay and Canariov (2023) from Umeå University, Sweden, provide an insightful and informative presentation of phenomenological research regarding professional role identity and experiences with generative AI in the work environment.

Their work found that “generative AI complements rather than replaces professionals, allowing them to focus on higher-level tasks,” yet they also highlighted that “precautions must be taken to build a framework that benefits people without destroying their values and cultures” (p. 48). Farazouli and fellow researchers from Stockholm University, Sweden, conducted an exploratory study of how AI chatbots impacted university teachers' assessment practices, by using a post-phenomenological approach (2023). They primarily used focus groups instead of longer and more direct personal interviews with faculty. They identified key findings such as "participants [faculty] seemed to adopt a more critical stance when assessing student texts. In several cases, the participants downgraded previous student responses as they did not accept mistakes such as repetitions, short answers and lack of engagement with course literature" (Farazouli et al, 2023, p. 9).

A phenological study, conducted by researchers at University of Technology and Applied Sciences – Ibra, Oman, and using in-depth interviews, identified key themes in both adoption and prevention of adoption of AI in education by faculty (Al-Mughairi et al., 2024). The themes identified as increasing AI adoption were: 1) interest in exploring innovative educational technologies, 2) personalization potential of AI in teaching and learning, 3) time-saving capabilities, and 4) professional development. Themes identified that inhibited AI adoption were: 1) reliability and accuracy concerns, 2) reduced human interaction, 3) privacy and data security, 4) lack of institutional support, and 5) fears of possible overreliance on ChatGPT (generative AI).

## Materials and Methods

This exploratory phenomenological qualitative research study (as derived from interpretivism, derived from constructivism) used multiple methods of traditional data collection (researchers' field notes, procedural observations, and in-depth open-ended, semi-structured, conversational interviews (Creswell, 2007). A phenomenological study was chosen in that it is based on the idea that unique experiences can be understood by obtaining proper, multi-layered description from those involved and is an effective method in the study of phenomena, social situations, and processes (Creswell, 2007; Yilmaz, 2013). Additionally, a phenomenological approach was used in that, as expressed and argued for by University of Louisiana at Lafayette educational researcher, Sebnem Cilesiz, it "provides a suitable framework for research on experiences in educational technology and can advance the field by complementing and unifying existing research" (2011, p. 488). IRB approval was obtained for all portions of this research from the American University of Armenia's Institutional Review Board.

Semi-structured interviews were chosen to provide deep insights, for maximum flexibility with the interviewees, and in that it is generally the most recommended for qualitative research (Adhabi & Anozie, 2017). These face-to-face, in-depth interviews were conducted with five different faculty members (from a pool of 19 available faculty members) during the final month of a university's spring semester. This number of a sample size have been shown to be effective when used with phenological interviews (Creswell, 1998, 2007; Santos et al., 2026). Emails were sent to all 19 available faculty with 7 responding, 2 stating that they strongly oppose AI and did not want to participate, and 5 stating that they would like to share their experiences dealing with AI in education. This convenience sample represented both male/female, full-time/part-time, and an age distribution from 20s to 60s. The homogeneous characteristics of the sample were that they were all actively teaching instructors during the Spring 2024 semester, at a university dealing with the new freely available generative AI technology (example ChatGPT). The sample size of five, was deemed appropriate in that this was a phenomenological research study requiring deep interviews and interaction (Creswell, 2007; Polkinghorne, 1987).

The semi-structured interviews used 10 open-ended questions about faculty's specific experiences with AI in teaching and learning, along with additional probing when needed. Interviews were one hour long with an opportunity for follow-on discussions if needed. All interviews were captured via audio means, and transcoded into transcripts for coding of themes and analysis. Independent inductive coding was accomplished by the two researchers with themes emerging from the data itself. A thematic analysis approach was used as outlined by Braun and Clarke (2017), with general categories dealing with faculty concerns, experiences, and feelings identified with resulting themes organized within each category (open coding, axial coding, and selective coding stages utilized). Multiple close listening of audio and reading of transcripts and field notes were used to help ensure reliability as well as multiple coding passes and deep discussions to reach inter-coder agreement.

## Results

**RQ1** What are the perceived experiences of faculty with their use of generative AI in the teaching and learning process?

Faculty shared that they had complex, often conflicting feeling about generative AI with their experiences in its use in teaching and learning. Sometimes these feelings were hopeful and positive, sometimes they were negative with lots of concerns. One dominant category that emerged, was faculties' concerns about academic integrity and student learning. The emerged theme was a preoccupation and *worry that AI would be used for cheating*. All faculty interviewed shared this concern with one instructor stating,

"I think basically, we just have to instill in them [students] from the get-go, a concept of integrity, honesty, and learning for one's use and advancement and improvement."

Two of the instructors expressed how they have been using the Turnitin AI text detection software and have had to interview students about it and in several cases penalize them for academic misconduct.

Another immerged and somewhat overlapping theme within this category was that faculty were now *using more in class assessments to ensure holding students' accountable for learning (without AI).* As an example, one instructor expressed that they had to abandon using online forums (message boards) because of a feeling that AI was being used both to write an initial response as well as to write a feedback response by the students. The instructor stated that they now do in-class quizzes to ensure that students keep up with the reading to know the material being discussed in class.

Another instructor directly expressed that,

“What I do in my class is to design assignments so that they cannot use AI."

Other instructors expressed using more project/problem-based learning and some in class writing to help address the issue.

An additional theme was a *worry that all writing will become the same if everyone is using AI to write.* This issue was summoned up by one faculty who stated,

> …my other worry, and this is what I've kind of communicated to the students, is that if you're all relying on the same tools to improve your writing, eventually everyone's writing is probably just going to sound the same. Because it's just going to collapse into one way of doing it. And that again, that's something that at least for the creativity side of English and communications, we're trying to foster individual voices and confidence in those voices.

Similarly, another faculty member expressed,

> If we all get started with trusting our creativity and intelligence to AI, imagine we all start doing that, what would be the result? …Then it will be cloning. I mean, in that, let's say, negative utopia. Right? It's a negative utopia.

When asked about using AI to provide feedback for students, the theme emerged of a *belief that using AI to provide feedback would be inappropriate*. One instructor felt that it would be a “disservice” to have the AI provide the feedback and additionally felt that they could do it much better. Another faculty presented a conflicted response by saying,

“I feel a bit uncomfortable; I haven't done so. But I believe I should do it, actually.”

Yet, even with these academic dishonesty concerns, faculty still expressed some level of positivity towards the new technology, with an emerged theme of an *overall feeling that AI is generally a good thing and is the future.* One faculty even stated "Maybe this [AI] is a move in the right direction. We can get rid of a lot of redundancies."

**RQ2** How do faculty describe their motivations and/or hesitations (worries/concerns) about using generative AI in teaching and learning?

Most faculty did not describe motivations for using AI and instead focused on their hesitations. A primary category for this was in quality concerns. The first emerged theme here is a *belief that AI is not creative or that those using AI are not creative*. One faculty member expressed this by stating,

> We don't want you [students] to write to a format. We don't want you to just give us information. We want you to challenge ideas. We want you to see things in new ways. We want you to be creative in your structures.

One faculty member expressed how they were angry, sad, and let down when several of the students admitted to using AI for a creative writing assignment, even when it was low stakes and the purpose of the assignment was to use their creativity.

Another quality concern hesitation theme was *a belief that AI isn't very effective or high quality*. One instructor, in talking about generative AI, plainly stated,

“I think it's functional but it's not necessarily of the highest quality.”

Another instructor explained that in their experience, AI results were less than satisfactory:

“Yeah, it would be hard because once I get stuff, it will seem a bit shallow, probably, not deep to me.“ However this same instructor also acknowledged that students were able to get “good grades” on writing assignments before Turnitin was used to help identify those inappropriately using the AI.

In yet another example of confliction, one instructor shared that going forward AI is needed in that they have a feeling that

> …some assessment methods probably do need to be overhauled in the long term. I mean, essays are great, but what skills are we teaching through them? How useful is that when you kind of move out of academia? Not everybody is going to be writing essays forever. I have to, but I chose that life, and that's a few people that will come out of an English program. What are they actually going to need to do? And I think that's something that, I'm still not 100% clear on… now it's a case of looking well within this sort of framework of assessment.

**RQ3** How do faculty describe any impediments in the use of generative AI in the teaching and learning process?

A major category that emerged dealt with faculties’ lack of training, understanding, and/or experience with AI. A primary theme here was that faculty did *not understand how AI could be used in different ways to enhance assignments while still focusing on the Student Learning Outcome (SLO)*. One instructor shared,

> I don't think I know enough, perhaps, or feel comfortable enough to do that yet. But I know it's the trend… I feel like I have such a lack of knowledge and everyone is probably 100 steps ahead of me.

Several of the other instructors also expressed the theme of *not using AI more due to lack of knowledge, direct support, and encouragement from a close trusted source*. Expressions of “ignorance” of the topic, not being “tech savvy,” and just not realizing the pedagogical possibilities. One faculty member did however express their efforts,

> The assignment that they ended up having was just get ChatGPT to do the thing that they were supposed to do, and then they had to write about getting ChatGPT to do the thing and what the limitations were there, how much babying it needed, whether you could trust it.

Faculty were directly asked if they were “aware that there are university resources to help faculty with AI.” All faculty stated that they were indeed aware that such resources existed (self-seeking improvement category with awareness of resources as a theme). They were then all asked if they had attended any of the multiple training sessions offered on using AI for teaching and learning. All faculty stated that they had not attended any of the training. When asked why they did not attend any of the training, faculty expressed many multi-layered reasons.

One instructor felt that even though they had used AI a little bit, they didn’t have enough knowledge to properly learn from the training. They felt intimidated and didn’t think they were ready to attend training yet. Another instructor, going along with the theme *not using AI more due to lack of knowledge, direct support, and encouragement from a close trusted source* expressed that they didn’t think it was important enough and that “I was still working it out for myself, and working out my own take on it… I don't think there's like a consensus, even within the department, as to what our AI policy should be."

Similarly, another instructor expressed not even being sure if it was appropriate to use AI in the classroom at all. They additionally stated that they had a bit of "arrogance" and felt that they already knew everything that was important about AI. Since they were highly educated within their field, they felt they should be able to figure everything out on their own.

All instructors also shared that the only generative AI they had used was the free version of ChatGPT. They additionally shared that the type of training they would prefer was hands-on basic workshops, asynchronous short online courses, and training within regular program meetings.

Finally, when asked what their overall feeling (attitudes) were regarding their experiences with AI in education (and limited to just a one sentence response), faculty expressed reservations but optimism at the same time: "AI can be a great help, but must be used wisely." "A lot of potential, but we need to be prepared to change if we want the students to." "Don't get carried away with it or let it become your conscious." "Ambiguously excited maybe." "AI is helpful, we need to use it wisely and teach our students to use it the same way."

**Table 1.** Categories and Themes Emerged from Phenomenological Interviews

*Categories and Themes Emerged from Phonological Interviews*

| CATEGORIES | THEMES |
|---|---|
| Concerns About Academic Integrity and Student Learning | Worried that AI will be used for Cheating |
| | Using more in class assessments to ensure holding students' accountable for learning (without AI) |
| | Worry that all writing will become the same if everyone is using AI to write. |
| | Belief that using AI to provide feedback would be inappropriate |
| Quality Concerns | Belief that AI is not creative or that those using AI are not creative. |
| | Belief that AI isn't very effective or high quality |
| Lack of Training, Understanding, and/or Experience | Not understanding how AI could be used in different ways to enhance the assignment while still focusing on the SLO. |
| | Have only experienced the free version of AI (GPT3.5 at the time). *Not the enhanced model with multi-modality capabilities*. |
| | Not using AI more due to lack of knowledge, direct support, and encouragement from a close trusted source. |
| Positive Feeling/Attitude | Overall feeling that AI is generally a good thing and is the future |
| Self-Seeking Improvement | Aware that there are university resources/training to help faculty with AI |

## Discussion

This phenomenological study identified that faculty interviewed had experiences with a great deal of issues and concerns while also producing optimism to its potential use. This potential enhanced use of AI in teaching and learning cannot be reached if identified factors such as encouragement, support, and effective training/resources are not fully developed and made available to faculty. Additionally, leadership and professional developers must help faculty understand that generative AI is much more than just a tool for cheating, in that many of the faculty seem to primary focus on that aspect. The Word Economic Forum and other researchers have expressed the explicit need for leadership to actively promote the balanced and purposeful use of AI and AI

Literacy development (Anders, 2024; Crawford et al., 2023; Partovi & Yongpradit, 2024; World Economic Forum, 2026). Although training and other resources were made available to the faculty interviewed, leadership needs to encourage its use and remind faculty of its availability while also working to cater resources to best fit/benefit faculty utilization.

Most of the faculty interviewed did not fully understand that AI could be integrated in multiple ways to directly help faculty prepare for their courses, assist in the actual presentation of content, be part of assignments/assessments, as well as help students learn the content. A better understanding of how to use generative AI within the overall teaching and learning process seems to also be needed to help faculties' pedagogical capabilities and to aid with course alignment. Multiple research in this area has shown the positive benefits that can be obtained when using generative AI in this way to enhance student-centered hands-on learning (Acun & Acun, 2023; Anders, 2025; Salinas- Navarro et al., 2024; Yilmaz & Yilmaz, 2023). Special presentations, conferences, workshops, as well as online courses are highly recommended along with special degree-program meetings (with leadership involvement) to better and more directly discuss the issues involved. Ongoing research from Carnegie Mellon University has also shown that learns prefer this type of AI developmental instruction in the form of experiential learning (Warrier et al., 2025). Finally, understanding that both AI literacy (based on a strong AI literacy framework such as the *Compass Framework for AI Literacy Integration into Higher Education*), is needed for all, and that special degree specific AI skills are needed for all faculty and students, as well as staff, is important in order to help everyone succeed with AI (Anders, 2025; Chukhlomin, 2026).

**Acknowledgments**

We wish to thank the English and Communication Program at the American University of Armenia for letting us conduct this study to help the body of knowledge in learning science dealing with generative artificial intelligence in education.

**Statements and Declarations**

Not applicable

**Ethical Considerations**

IRB approval was obtained for all portions of this research from the American University of Armenia's Institutional Review Board.

**Consent to Participate**

Informed consent was given via verbal means in the presence of both researchers, by each of the participants, as per our University's IRB guidelines.

**Declaration of Conflicting Interest**

The author(s) declared no potential conflicts of interest with respect to the research, authorship, and/or publication of this article

**Funding Statement**
The author(s) received no financial support for the research, authorship, and/or publication of this article.

**Data Availability**

Data has not been made publicly available to ensure the anonymity of the study participants. Redacted interview content can be made available upon request.